\documentclass[preprint,3p]{elsarticle}

\def \eps{\epsilon}
\def \pipj{(p_i\cdot p_j)}
\def \piq{(p_i\cdot q)}
\def \pjq{(p_j\cdot q)}
\def \ai{\alpha_i}
\def \aj{\alpha_j}
\def \Li{{\rm Li}}

\begin{document}

\title{The singular behavior of one-loop massive QCD amplitudes with
  one external soft gluon}

\author[Aahen]{Isabella Bierenbaum}
\author[Aahen]{Micha\l{}  Czakon}
\author[CERN]{Alexander Mitov}

\address[Aahen]{Institut f\"ur Theoretische Teilchenphysik und Kosmologie,
RWTH Aachen University, D-52056 Aachen, Germany}
\address[CERN]{Theory Division, CERN, CH-1211 Geneva 23, Switzerland}

\date{\today}

\cortext[thanks]{Preprint numbers: TTK-11-28, CERN-PH-TH/2011-181}

\begin{abstract}
We calculate the one-loop correction to the soft-gluon current with
massive fermions. This current is process independent and controls the
singular behavior of one-loop massive QCD amplitudes in the limit when
one external gluon becomes soft. The result derived in this work is
the last missing process-independent ingredient needed for numerical
evaluation of observables with massive fermions at hadron colliders at
the next-to-next-to-leading order.
\end{abstract}
\maketitle


\section{Introduction}

The main obstacle for the numerical evaluation of collider observables
at higher perturbative orders is the presence of infrared (IR) (soft
and collinear) divergences in parton level calculations. These
divergences cancel in observables, but need to be regularized in all
intermediate calculations by introducing an appropriate parameter
(typically dimensionally). It is the need to keep track of such a
regularization parameter that prevents the {\it ab initio} application
of straightforward methods for numerical integration.

At the next-to-leading order (NLO), this complication can be evaded
within the so-called subtraction method. Its basic idea is simple:
first, one utilizes the universality and factorization property of IR
singularities to construct an approximation to the corresponding real
emission partonic amplitude. This approximation is simple enough to
allow the analytic extraction,  and eventually cancellation, of IR
singularities. Second, one explores the fact that the difference of
the full amplitude and its approximation is IR finite and therefore
can be integrated numerically in a straightforward way. This approach,
in effect, splits the task of performing a complicated divergent
integration in two: first, a divergent integration of a simpler
quantity and, second, a complicated, but finite numerical
integration. The subtraction approach can be applied to processes with
massless and massive fermions. Several subtraction schemes have been
proposed
\cite{Giele:1991vf,Giele:1993dj,Frixione:1995ms,Catani:1996vz,Frixione:1997np,Catani:2002hc}
and have been successfully used in a large number of applications.

To construct a subtraction scheme at next-to-next-to-leading order
(NNLO) one needs to know, among others, the limiting behavior of
one-loop amplitudes when one of the external {\it on-shell} partons -
a gluon - becomes soft. For the case of massless fermions (like
massless QED or QCD) this problem has been studied in specific cases
in Refs.~\cite{Bern:1998sc,Bern:1999ry} and later in
Ref.~\cite{Catani:2000pi}. The goal of the present work is to
generalize the process-independent approach of Catani and Grazzini
\cite{Catani:2000pi} to the case of massive fermions. With the result
of one of the authors for the treatment of double real radiation
\cite{Czakon:2010td,Czakon:2011ve}, the result derived in this paper
represents the last missing ingredient for the construction of NNLO
observables with massive fermions. Applications are top and bottom
(and charm) production at hadron colliders, deep inelastic scattering
and processes at lepton colliders.

This paper is organized as follows: In Section \ref{sec:factorization}
we present the factorization of one-loop amplitudes in the soft limit
and introduce the one-loop soft-gluon current. There we also present
the derivation of the one-loop soft-gluon current in terms of a set of
scalar integrals. In Sections \ref{sec:case1}, \ref{sec:case2} and
\ref{sec:case3} we present the explicit result for the UV
unrenormalized one-loop soft-gluon current for all phenomenologically
relevant kinematical configurations. In Section \ref{sec:checks} we
discuss the properties of the results and explain a number of checks
we have performed. In Section \ref{sec:squared} we derive the soft
limit of squared matrix elements as needed in specific
applications. In Section \ref{sec:UVren} we describe the UV
renormalization of the bare one-loop soft-gluon current followed by a
summary. We have added a number of appendices containing most of the
technical details. The evaluation of all scalar integrals is detailed
in \ref{sec:app-integrals}. In \ref{sec:app-small-mass} and
\ref{sec:app-poles} we independently derive two limiting results for
the UV renormalized one-loop soft current: its small-mass limit and
its pole terms, respectively. Finally, in \ref{sec:analytical-cont},
we discuss the analytical continuation of the bare one-loop soft-gluon
current evaluated in different kinematical configurations.


\section{Amplitude factorization in the soft limit}\label{sec:factorization}

Consider the amplitude $M_a(n+1;q)$ for producing $n+1$ on-shell
partons.
\footnote{Unless we state otherwise, we do not make a distinction
  between initial and final state partons.}
Let at least one final state parton be a gluon, and let $a=1,\dots ,
N_c^2$ be its color index and $q$ its momentum $q^2=0$. It is useful
to think of  $M_a(n+1;q)$ as a wide-angle scattering amplitude,
i.e. all kinematical invariants formed from its external momenta are
large. The structure of such amplitudes is very well understood
through at least two-loops
\cite{Sen:1982bt,Sterman:2002qn,Aybat:2006mz,Mitov:2006xs,Gardi:2009qi,
  Kidonakis:2009ev,Becher:2009qa,Mitov:2009sv,Becher:2009kw,Czakon:2009zw,
  Ferroglia:2009ep,Ferroglia:2009ii,Dixon:2009ur,Mitov:2010xw,Gardi:2010rn,
  Mitov:2010rp} in both the massive and the massless cases. We are
next interested in the limit when the external gluon becomes soft
$q\to 0$, or more precisely its momentum scales as: 
\begin{equation}
q \to \lambda q , ~\lambda \to 0 \, .
\label{eq:softlimit}
\end{equation}
Clearly, in the limit (\ref{eq:softlimit}), $M_a(n+1)$ is not a
wide-angle scattering amplitude anymore. Instead, it satisfies the
following factorization property:
\begin{equation}
M_a(n+1;q) = J_a(q) M(n) + {\cal O}(\lambda) \, .
\label{eq:M-fact}
\end{equation}
In the above equation, the amplitude $M(n)$ is the wide-angle
scattering amplitude obtained from $M_a(n+1;q)$ by removing the
external gluon with momentum $q$, and $J_a(q)$ is the
process-independent soft-gluon (eikonal) current whose derivation is
the main goal of this work. We have adopted a concise notation in
Eq.~(\ref{eq:M-fact}), but have made explicit the dependence on $q, a$
and $n$.

Each one of the factors in Eq.~(\ref{eq:M-fact}) depends on the
dimensional regularization parameter $\epsilon =(4-d)/2$ and has a
loop expansion in powers of the strong coupling constant through one
loop:
\begin{eqnarray}
J_a(q) &=& g_S\mu^\epsilon \left( J_a^{ (0)}(q) +  J_a^{ (1)}(q) + \dots \right) \, , \nonumber\\
M(n) &=&  M^{(0)}(n) +  M^{(1)}(n) + \dots \, ,\nonumber\\ 
M_a(n+1;q) &=& M_a^{(0)}(n+1;q) + M_a^{(1)}(n+1;q) + \dots \, ,
\label{eq:expansion}
\end{eqnarray}
where the dots stand for terms at higher orders in $\alpha_S$. 
The notation adopted in Eq.~(\ref{eq:expansion}) does not make
explicit the powers of the strong coupling $\alpha_S$. For example,
$J_a^{ (0)}$ denotes the leading order result for the soft current
(given explicitly in Eq.~(\ref{eq:J0}) below), $J_a^{ (1)}$ stands for
its next-to-leading order in $\alpha_S$, and so on. The reason for
choosing this notation is that the leading order amplitude
$M^{(0)}(n)$ contains a process dependent power of the strong coupling
constant. Thus, our notation reflects the only relevant information: 
the power of the strong coupling relative to the leading
order amplitude  $M^{(0)}(n)$.

Our considerations apply for both bare and UV renormalized
amplitudes. For now we consider bare amplitudes and will return to the
UV renormalization in Section \ref{sec:UVren}.

Both in the massive ($p_i^2>0$) and massless ($p_i^2=0$) cases the
tree level soft-gluon current reads:
\begin{equation}
 J_a^{\mu (0)}(q) = \sum_{i=1}^n T_i^a {p_i^\mu \over p_i\cdot q} \equiv \sum_{i=1}^n T_i^a e_i^\mu \, ,
 \label{eq:J0}
\end{equation}
where $J_a^{ (n)}(q)\equiv \varepsilon^\mu(q) J_a^{\mu
  (n)}(q)$. Throughout we follow the conventions of
Ref.~\cite{Catani:2000pi} for the signs of color generators. The one-loop UV un-renormalized soft-gluon current $
J_a^{\mu (1)}(q)$ reads:
\begin{eqnarray}
J_a^{\mu (1)}(q) = if_{abc}\sum_{i\neq j =1}^n T_i^bT_j^c \left(e_i^\mu - e_j^\mu\right) g_{ij}^{(1)}(\epsilon,q,p_i,p_j)\, .
\label{eq:J1}
\end{eqnarray}

For the calculation of the one-loop soft-gluon current in
Eq.~(\ref{eq:J1}) we follow the strategy developed in
Ref.~\cite{Catani:2000pi}. The approach consists of the evaluation of
all one-loop diagrams connecting (on-shell) external legs and
attaching a real gluon to either the external legs or the gluon
propagator (the virtual gluon). The calculation is performed in the
eikonal approximation, i.e. the real and virtual gluons are treated as
being of similar magnitude, and energy-momentum conservation is
enforced.  

Following the terminology introduced in Ref.~\cite{Catani:2000pi}, we
split the results in $1P$ and $2P$ contributions. The $1P$
contributions are defined as the ones that depend on a single external
hard momentum $p_i$, as opposed to the $2P$ contributions that involve
two hard momenta $p_i$ and $p_j$. In the following we calculate the
$2P$ contributions and show that they are separately conserved; then,
adapting the arguments given in Ref.~\cite{Catani:2000pi} one can show
that the $1P$ terms do not contribute to the soft current.

Our starting point for the calculation of the $2P$ contribution $ J_{a
  (2P)}^{\mu (1)}$ to the one-loop soft-gluon current is the sum of
the three diagrams $4(a,b,c)$ given in Ref.~\cite{Catani:2000pi}.
\footnote{Since, for these diagrams, at the integrand level the
  eikonal approximation is identical in the massive and the massless
  cases, we can simply use the sum of the expressions given in
  Eq.~(46,47) of Ref.~\cite{Catani:2000pi}. We have verified the
  agreement.}
We neglect all scaleless integrals. The sum of diagrams is gauge
invariant as also explained in \cite{Catani:2000pi}. The term $\sim
k\cdot \varepsilon (q)$ is reduced to scalar integrals. The reduction
differs from the one in Ref.~\cite{Catani:2000pi} since in the case at
hand, it produces terms that explicitly depend on the masses
$m_{i,j}^2$. Applying partial fractioning and omitting scaleless
integrals, we arrive at the following expression for the function
$g^{(1)}_{ij}$ in Eq.~(\ref{eq:J1}):
\begin{eqnarray}
g^{(1)}_{ij} &=& a_S^b\mu^{2\eps}~ {p_i\cdot p_j \over m_i^2(p_j\cdot q)^2-2(p_i\cdot p_j)(p_i\cdot q)(p_j\cdot q) + m_j^2(p_i\cdot q)^2} \nonumber\\
&\times& \Bigg\{ (p_i\cdot q)(p_j\cdot q)\left[(p_j\cdot q)M_1 + (p_i\cdot q)\hat{M}_1\right] \nonumber\\
&&+ {1\over 2}(p_j\cdot q)\left[(p_i\cdot p_j)(p_i\cdot q) - m_i^2(p_j\cdot q)\right] M_2 + {1\over 2}(p_i\cdot q)\left[(p_i\cdot p_j)(p_j\cdot q) - m_j^2(p_i\cdot q)\right] \hat{M}_2 \nonumber\\  
&&+ \left[ (p_i\cdot p_j)(p_i\cdot q)(p_j\cdot q) - m_i^2(p_j\cdot q)^2 - m_j^2(p_i\cdot q)^2\right]{(p_i\cdot q)(p_j\cdot q)\over p_i\cdot p_j} M_3 \Bigg\} \, .
\label{eq:gij-masters}
\end{eqnarray}
The bare coupling $a_S^b=\alpha_s^bS_\eps/(2\pi)$ with $S_\eps =
(4\pi)^\eps\exp(-\eps \gamma_E)$, $\hat{M}_k \equiv
M_k(p_i\leftrightarrow p_j),~k=1,2,3$ and the integrals $M_{1,2,3}$
can be found in \ref{sec:app-integrals}. Noting that $\hat{M}_3 =
M_3$, it is apparent that $g^{(1)}_{ij}=g^{(1)}_{ji}$. From
Eqs.~(\ref{eq:J0},\ref{eq:J1}) it is evident that the massive
soft-gluon current is conserved through one loop. This follows from
color conservation (as explained in Ref.~\cite{Catani:2000pi}) and the
identity $q\cdot e_i=1$.

Next we present our main result, namely, the explicit expression for
the function $g^{(1)}_{ij}$. There are three kinematical regions for
which this function needs to be computed. In all three cases we
consider $p_i^2=m_i^2 > 0$ and take $p_i$, as well as the momentum $q$
of the soft-gluon, to be in the final state. Thus, the three
kinematical configurations are defined as:
\begin{enumerate}
\item $p_j^2 = 0,~p_j$~incoming,
\item $p_j^2 = 0,~p_j$~outgoing,
\item $p_j^2 =m_j^2 > 0,~p_j$~outgoing.
\end{enumerate}

We do not have in mind phenomenological applications with massive
quarks in the initial state, but for completeness, have calculated and
presented below all required ingredients for such applications as well.


\subsection{Case 1}\label{sec:case1}

We have evaluated this kinematical configuration directly, as
described above, by substituting the explicit results
Eqs.~(\ref{eq:M1}, \ref{eq:M2-exact}, \ref{eq:M3-1mass}) for the
scalar integrals into Eq.~(\ref{eq:gij-masters})  and then expanding
in epsilon to the desired depth. The result for the un-renormalized
one-loop soft current reads:
\begin{eqnarray}
g^{(1)}_{ij}({\it Case~1}) = R_{ij}^{[C1]} + i\pi I_{ij}^{[C1]} 
\equiv ~a_S^b \left({2\pipj \mu^2\over 2\piq 2\pjq}\right)^\eps \sum_{n=-2}^2 \eps^n \left(  R^{(n)[C1]}_{ij} + i\pi I^{(n)[C1]}_{ij} \right) \, ,
\label{eq:gij-explicitCase1}
\end{eqnarray}
The bare coupling $a_S^b$ is introduced in Eq.~(\ref{eq:gij-masters}) and:
\begin{eqnarray}
I^{(-2)[C1]}_{ij} &=& 0 \, , \label{eq:Result-Case-1}\\
I^{(-1)[C1]}_{ij} &=& -{1\over 2} \, , \nonumber\\
R_S\, I^{(0)[C1]}_{ij} &=& 2m_i^2 \pjq \ln\left({\ai \over 2}\right)\, , \nonumber\\
R_S\, I^{(1)[C1]}_{ij} &=& 4\left[\pipj \piq - m_i^2 \pjq\right] \Li_2\left(1- {\ai \over 2} \right)  + m_i^2 \pjq \ln^2\left({\ai \over 2}\right) 
\nonumber\\
&& + \pi^2 {-2 \pipj \piq + m_i^2 \pjq\over 2} \, , \nonumber\\
R_S\, I^{(2)[C1]}_{ij} &=& 4 \left[\pipj \piq - m_i^2 \pjq\right] \left[\Li_3\left(1- {\ai \over 2} \right) + \Li_3\left({\ai \over 2}\right) \right] - \zeta_3 {40 \pipj \piq - 26 m_i^2 \pjq \over 3} \nonumber\\
&& + 2\left[\pipj \piq - m_i^2 \pjq\right] \ln\left(1- {\ai \over 2} \right) \ln^2\left({\ai \over 2}\right) +{m_i^2 \pjq\over 3}\ln^3\left({\ai \over 2}\right) \nonumber\\
&& + \ln\left({\ai \over 2}\right) \left( \pi^2 {-4 \pipj \piq + m_i^2 \pjq\over 6} + 4 \left[\pipj \piq - m_i^2 \pjq\right] \Li_2\left(1- {\ai \over 2} \right)\right)\nonumber\\ 
&& \nonumber\\
R^{(-2)[C1]}_{ij} &=& -{1\over 2} \, , \nonumber\\
R^{(-1)[C1]}_{ij} &=& 0  \, , \nonumber\\
R_S\, R^{(0)[C1]}_{ij} &=&  m_i^2 \pjq \ln^2\left({\ai \over 2}\right) -\pi^2 {5(2 \pipj \piq - m_i^2 \pjq) \over 6}\, , \nonumber\\
R_S\, R^{(1)[C1]}_{ij} &=& 4 \left[\pipj \piq - m_i^2 \pjq\right] \Li_3\left({\ai \over 2}\right) - \zeta_3 {4\left[7 \pipj \piq - 5 m_i^2 \pjq\right]\over 3} \nonumber\\
&& + 2 \left[\pipj \piq - m_i^2 \pjq\right] \ln\left(1- {\ai \over 2} \right) \ln^2\left({\ai \over 2}\right) \nonumber\\
&&+ \ln\left({\ai \over 2}\right) \left( \pi^2{-2 \pipj \piq - 5 m_i^2 \pjq\over 3 } + 4 \left[\pipj \piq - m_i^2 \pjq\right] \Li_2\left(1- {\ai \over 2} \right) \right)\, ,\nonumber\\
R_S\, R^{(2)[C1]}_{ij} &=& - 4 \left[\pipj \piq - m_i^2 \pjq\right] \left[\Li_4\left(1 - {2\over \ai}\right) + \Li_4\left(1- {\ai \over 2} \right) - \Li_4\left({\ai \over 2}\right)\right]\nonumber\\
&& +\pi^4 {458 \pipj \piq - 213 m_i^2 \pjq\over 720 } \nonumber\\
&& + \ln\left({\ai \over 2}\right) \left( 4 \left[\pipj \piq - m_i^2 \pjq\right] \Li_3\left(1- {\ai \over 2} \right) - 2\zeta_3 \left[2 \pipj \piq - m_i^2 \pjq\right] \right)\nonumber\\
&& + \pi^2{-4 \pipj \piq - m_i^2 \pjq\over 12} \ln^2\left({\ai \over 2}\right) + 2{\pipj \piq - m_i^2 \pjq\over 3} \ln\left(1- {\ai \over 2} \right) \ln^3\left({\ai \over 2}\right)\nonumber\\
&& + {-2 \pipj \piq + 3 m_i^2 \pjq\over 12} \ln^4\left({\ai \over 2}\right) - \pi^2 {14\over 3}\left[\pipj \piq - m_i^2 \pjq\right] \Li_2\left(1- {\ai \over 2} \right) \, . \nonumber
\end{eqnarray}
We find it convenient to express the result through the variables
$R_S$ and $\ai$ defined as:
\begin{equation}
R_S = 4\left[m_i^2 \pjq - 2 \pipj \piq \right] ~,~ \ai = {m_i^2(p_j\cdot q)\over (p_i\cdot q) (p_i\cdot p_j) } ~,~ \aj = {m_j^2(p_i\cdot q)\over (p_j\cdot q) (p_i\cdot p_j) } \, .
\label{eq:def-ai-aj}
\end{equation}
We introduce the variable $\aj$ for later use. The result in
Eq.~(\ref{eq:Result-Case-1}) is also available in electronic form.


\subsection{Case 2}\label{sec:case2}

To derive this case we have performed an analytical continuation of
the result from $Case~1$. When  $p_j^2=0$ the continuation amounts to
exchanging $p_j\to - p_j$ (see \ref{sec:analytical-cont}). It is easy
to see that Eq.~(\ref {eq:Result-Case-1}) remains unchanged under this
transformation, i.e. the result in $Case~2$ is identical to that in
$Case~1$.


\subsection{Case 3}\label{sec:case3}

We calculate the result for {\it Case 3} up to and including
terms of ${\cal O}(\epsilon)$. We note that the ${\cal O}(\epsilon^2)$
term contributes only, if multiplied by a term $\sim 1/\epsilon^2$
originating from the phase-space integration. Such a leading pole may
only be due to the emission of soft and collinear radiation. Both
partons $i$ and $j$ beeing massive, collinear singularities
are regularized and do not lead to poles in $\epsilon$.

We compute the result for the one-loop soft-gluon current in this
kinematical configuration directly, as expansion in
$\epsilon$. Details of the calculation can be found in
\ref{sec:app-integrals}. For completeness, and with more formal
applications in mind, we have also calculated all integrals in the
spacelike region, where the massive momentum $p_j$ is incoming; see
\ref{sec:app-integrals}. The relation between the results in the two
kinematical configurations is discussed in \ref{sec:analytical-cont}.

The explicit result for the soft-gluon current in the kinematics of
{\it Case 3} reads:
\begin{eqnarray}
g^{(1)}_{ij}({\it Case~3}) = R_{ij}^{[C3]} + i\pi I_{ij}^{[C3]} \equiv ~a_S^b  \left({2\pipj \mu^2\over 2\piq 2\pjq}\right)^\eps \sum_{n=-2}^1 \eps^n \left( R^{(n)[C3]}_{ij} +i\pi I^{(n)[C3]}_{ij}  \right)\, .
\label{eq:gij-explicitCase3}
\end{eqnarray}
and:
\begin{eqnarray}
I^{(-2)[C3]}_{ij} &=& 0 \, , \label{eq:Result-Case-3}\\
I^{(-1)[C3]}_{ij} &=& -1+{1\over 2v} \, , \nonumber\\
I^{(0)[C3]}_{ij} &=& 
{\ln(v)\over v} + {\ln(x)\over 2 v} +\left(1 + {1\over 2 v}\right)\ln(1 + x^2)  \nonumber\\ 
&& + {1\over Q_S}\left[-4 {m_j^2 \piq^2 - m_i^2 \pjq^2\over v} \ln\left({\ai\over \aj}\right) + 16 \pipj \piq \pjq \ln(x) \right]\, , \nonumber\\
&& \nonumber\\
I^{(1)[C3]}_{ij} &=& 
\frac{1}{v}\left(
\frac{1}{16} \ln ^2\left(\frac{\ai}{\aj}\right)+\Li_2\left(x^2\right)+\ln (v) \left(\ln \left(x^2+1\right)+\ln (x)\right)+\ln ^2(v)+\frac{1}{4} \ln
   ^2\left(x^2+1\right)\right. \nonumber\\
&&+ \left. \frac{1}{2} \ln (x) \ln \left(x^2+1\right)+\frac{\ln ^2(x)}{4}-\frac{\pi ^2}{8}
\right) \nonumber\\
&&+
\frac{1}{Q_S}
\Big[
\pipj \piq \pjq \left(2 \ln ^2\left(\frac{\ai}{\aj}\right)-16 \ln \left(x^2+1\right) 
\ln (x)+8 \ln ^2(x)-\frac{8 \pi^2}{3}\right)\nonumber \\
&&
+\left(m_i^2 \pjq^2+m_j^2 \piq^2\right)\left(8 \ln^2\left(x^2+1\right)-\frac{4 \pi ^2}{3}\right)\nonumber \\
&&
-4 \left(m_j^2 \piq^2-m_i^2 \pjq^2\right) \frac{1}{v}
\left(2 \ln (v)+ \ln \left(x^2+1\right)+ \ln (x)\right) \ln \left(\frac{\ai}{\aj}\right)\nonumber \\
&&
+\left(m_j^2 \piq^2+m_i^2 \pjq^2-\pipj \piq \pjq\right)
\Big( \nonumber \\
&& 
32 \ln (2) \left(- \ln (\ai+v+1)- \ln (\aj+v+1)-2 \ln \left(x^2+1\right)- \ln (x)\right)+64 \ln^2(2) \nonumber \\
&&
+16 \ln \left(x^2+1\right) ( \ln (\ai+v+1)+ \ln (\aj+v+1)) 
+16 \ln (\ai+v+1) \ln (\aj+v+1)\nonumber \\
&&
+16 \ln (x) \left( \ln (-\aj+v+1)+ \ln (\aj+v-1)+2 \ln \left(x^2+1\right)\right)
-16 \ln ^2(x)
 \nonumber \\
&&
+ 8 \ln \left(\frac{\ai}{\aj}\right) ( \ln (\aj+v-1)- \ln (-\aj+v+1))
+4 \ln ^2\left(\frac{\ai}{\aj}\right) \nonumber \\
&&
+16 \Li_2\left(\frac{-v+\aj+1}{2 \aj}\right) 
+16 \Li_2\left(2-\frac{2 \aj}{-v+\aj+1}\right) 
-16 \Li_2\left(\frac{-v+\aj+1}{v+\aj+1}\right) \nonumber \\
&&
+16 \Li_2\left(\frac{v+\aj+1}{2 v+2}\right) 
-16 \Li_2\left(-\frac{(v-1) (v+\aj+1)}{(v+1) (-v+\aj+1)}\right) 
+16 \Li_2\left(\frac{2 \aj}{v+\aj+1}\right) 
\Big)
\Big]\, . \nonumber
\end{eqnarray}
\begin{eqnarray*}
R^{(-2)[C3]}_{ij} &=& -{1\over 2} \, , \nonumber\\
R^{(-1)[C3]}_{ij} &=& {1\over 2} \left(-1 + {1\over v}\right) \ln(x) +{1\over 2} \ln(1 + x^2)  \, , \nonumber\\
R^{(0)[C3]}_{ij} &=&  {1\over 2 v} \Li_2(x^2) + \pi^2 \left({19\over 24} - {7\over 12 v}\right) + {1\over v} \ln(v) \ln(x)   +{1\over 2}\left(1 +{1\over v}\right) \ln(x) \ln(1 + x^2) \nonumber\\
&&   - {1\over 4} \ln^2(1 + x^2)   + {1\over Q_S}\left[ \left(m_j^2 \piq^2 + m_i^2 \pjq^2\right) \ln^2\left({\ai\over \aj}\right) \right.\nonumber\\
&&\left. + 4\left(m_j^2 \piq^2 + m_i^2 \pjq^2\right) \ln^2(x) - 4{m_j^2 \piq^2 - m_i^2 \pjq^2\over v} \ln\left({\ai\over \aj}\right) \ln(x) \right] \, . \nonumber
\nonumber\\
R^{(1)[C3]}_{ij} &=& 
\frac{1}{v}
\left(
-\ln (v) \left(\ln (x) \ln \left(x^2+1\right)+\pi ^2\right)
+\frac{\ln ^3(x)}{12}+\frac{\zeta (3)}{2}
\right.
\nonumber\\
&& \left.
+\ln (x) 
\left(
\frac{1}{16} \ln ^2\left(\frac{\ai}{\aj}\right)
+\frac{\Li_2\left(x^2\right)}{2}
-\frac{3}{4} \ln ^2\left(x^2+1\right)
-\frac{5 \pi^2}{24}
\right)\right.
\nonumber\\
&&\left. 
-\left(\frac{\Li_2\left(x^2\right)}{2}+\frac{5 \pi ^2}{12}\right) \ln \left(x^2+1\right)
-\frac{1}{2} \left(2 \Li_3\left(1-x^2\right)+\Li_3\left(x^2\right)\right) 
\right)
\nonumber\\
&&
+\frac{1}{Q_S}
\Big[
\pipj\piq\pjq
\Big(
\frac{32 \ln ^3(x)}{3}-\frac{280 \zeta (3)}{3}-32 \ln \left(x^2+1\right) \ln ^2(x) \nonumber\\
&&
\nonumber\\
&&
+\ln \left(x^2+1\right) \left(36 \pi ^2-4 \ln ^2\left(\frac{\ai}{\aj}\right)\right) 
+\left(48 \ln^2\left(x^2+1\right)-\frac{40 \pi ^2}{3}\right) \ln (x)
\nonumber\\
&&
-\frac{88}{3} \ln ^3\left(x^2+1\right)
\Big)
\nonumber\\
&&
(m_j^2 \piq^2 + m_i^2 \pjq^2)
\Big(
\left(3 \ln \left(x^2+1\right)+\ln (x)\right) \ln ^2\left(\frac{\ai}{\aj}\right)
\nonumber\\
&&
+28 \ln ^3\left(x^2+1\right)
-44 \ln^2\left(x^2+1\right) \ln (x)
-\frac{70}{3} \pi ^2 \ln \left(x^2+1\right)
+28 \ln \left(x^2+1\right) \ln ^2(x)
\nonumber\\
&&
-\frac{28}{3} \ln ^3(x)
+\frac{2}{3} \pi ^2 \ln (x)
+\frac{224 \zeta (3)}{3}
\Big)
\nonumber\\
&&
-\frac{(m_j^2 \piq^2 - m_i^2 \pjq^2)}{v}
\ln \left(
\frac{\ai}{\aj}\right) 
\left(
4 \Li_2\left(x^2\right)
+4 \ln \left(x^2+1\right) \ln (x)
+8 \ln (v) \ln (x)
-\frac{14 \pi ^2}{3}
\right)
\nonumber\\
&&
+(m_i^2 \pjq^2 + m_j^2 \piq^2 - \pipj \piq \pjq)\Big(
\ln ^3\left(\frac{\ai}{\aj}\right)
- \ln ^2\left(\frac{\ai}{\aj}\right)2 \ln (v)
\nonumber\\
&&
+ \ln ^2\left(\frac{\ai}{\aj}\right)2 (\ln (\ai+v+1)+\ln (-\aj+v+1)+\ln (\aj+v-1)-3 \ln (2))
\nonumber\\
&&
\nonumber\\
&&
+\ln \left(\frac{\ai}{\aj}\right)
\Big(
2 \ln ^2(v) 
-12 \ln ^2(\ai+v+1)
+6 \ln^2(\aj+v+1)
-4 \ln ^2(x)
-12 \ln ^2\left(x^2+1\right)
\nonumber\\
&&
+28 \ln (2) \ln (\ai+v+1)
+4 \ln (v) (\ln (\ai+v+1)-2 \ln (\aj+v+1)+\ln (2))
-10 \ln ^2(2)
\nonumber\\
&&
-4 \ln (\ai+v+1) \ln (\aj+v+1)
-8 \ln (2) \ln (\aj+v+1)
\nonumber\\
&&
+8 (2 \ln (\ai+v+1)-\ln (-\aj+v+1)+\ln (\aj+v-1)+\ln (\aj+v+1)-3 \ln (2)) \ln (x)
\nonumber\\
&&
+24 (\ln (2)-\ln (\ai+v+1)) \ln \left(x^2+1\right)
+24 \ln (x) \ln \left(x^2+1\right)
-\frac{2}{3} \pi ^2
\Big)
\nonumber\\
&&
+\frac{32}{3} \ln ^3(\ai+v+1)
+12 \ln ^3(\aj+v+1)
-32 \ln (2) \ln ^2(\ai+v+1)
-8 \ln (v) \ln ^2(\aj+v+1)
\nonumber\\
&&
+4 \ln (\ai+v+1) \ln ^2(\aj+v+1)
-40 \ln (2) \ln ^2(\aj+v+1)
-8 \ln (v) \ln ^2(x)
-\frac{4}{3} \pi ^2 \ln (v)
\nonumber\\
&&
+8 (3 \ln (\ai+v+1)+\ln (-\aj+v+1)+\ln (\aj+v-1)-5 \ln (2)) \ln ^2(x)
\nonumber\\
&&
+40 (\ln (\ai+v+1)+\ln (\aj+v+1)-2 \ln (2)) \ln ^2\left(x^2+1\right)
+36 \ln ^2(2) \ln (\ai+v+1)
\nonumber\\
&&
+8 \ln (v) \ln (\ai+v+1) \ln \left(\frac{1}{2} (\aj+v+1)\right)
-4 \ln ^2(v) (\ln (\ai+v+1)-\ln (\aj+v+1))
\nonumber\\
&&
+\frac{4}{3} \pi ^2 (-3 \ln (\ai+v+1)-4 \ln (\aj+v+1)+7 \ln (2))
+8 \ln (2) \ln (v) \ln (\aj+v+1)
\nonumber\\
&&
-8 \ln (2) \ln (\ai+v+1) \ln (\aj+v+1)
+44 \ln ^2(2) \ln (\aj+v+1)
+4 \ln ^2(v) \ln (x)
\nonumber\\
&&
-24 \ln ^2(\ai+v+1) \ln (x)
-4 \ln ^2(\aj+v+1) \ln (x)
+8 \ln (v) (\ln (2)-\ln (\ai+v+1)) \ln (x)
\nonumber\\
&&
+56 \ln (2) \ln (\ai+v+1) \ln (x)
-8 \ln (\ai+v+1) \ln (\aj+v+1) \ln (x)
-36 \ln ^2(2) \ln (x)
\nonumber\\
&&
+16 \ln (2) \ln (\aj+v+1) \ln (x)
+32 \ln ^2(\ai+v+1) \ln \left(x^2+1\right)
+80 \ln ^2(2) \ln \left(x^2+1\right)
\nonumber\\
&&
+32 \ln ^2(\aj+v+1) \ln \left(x^2+1\right)
-80 \ln (2) \ln (\ai+v+1) \ln \left(x^2+1\right)
\nonumber\\
&&
+16 \ln (\ai+v+1) \ln (\aj+v+1) \ln \left(x^2+1\right)
-80 \ln (2) \ln (\aj+v+1) \ln \left(x^2+1\right)
\nonumber\\
&&
+16 (-4 \ln (\ai+v+1)-\ln (\aj+v+1)+5 \ln (2)) \ln (x) \ln \left(x^2+1\right)
-\frac{80 \ln ^3(2)}{3}
\nonumber\\
&&
+\left(8 \ln \left(\frac{\ai}{\aj}\right)+16 \ln (x)\right) \Li_2\left(\frac{1-v}{\aj}\right)
+\left(16 \ln (x)-8 \ln \left(\frac{\ai}{\aj}\right)\right) \Li_2\left(\frac{\aj}{v+1}\right)
\nonumber\\
&&
+\left(4 \ln \left(\frac{\ai}{\aj}\right)-8 \ln (\ai+v+1)-8 \ln (\aj+v+1)+24 \ln (x)-16 \ln \left(x^2+1\right)+4 \ln ^4 (2)\right) 
\nonumber\\
&&
\quad \times \left( 
\Li_2\left(\frac{v-1}{\aj}\right)-\Li_2\left(\frac{\aj}{\aj-v+1}\right)
\right)
\nonumber\\
&&
+\left(4 \ln \left(\frac{\ai}{\aj}\right)-8 \ln (\ai+v+1)-8 \ln (\aj+v+1)-8 \ln (x)-16 \ln \left(x^2+1\right)+4 \ln ^4 (2)\right) 
\nonumber\\
&&
\quad \times \left( 
\Li_2\left(-\frac{v+1}{\aj}\right)- \Li_2\left(\frac{\aj}{\aj+v+1}\right)
\right)
\nonumber\\
&&
+8( \ln (v)- \ln (\aj+v+1)+ \ln (2)) \Li_2\left(-\frac{(v-1) (\aj+v+1)}{(\aj-v+1) (v+1)}\right)
-16 \ln (x) \Li_2\left(x^2\right)
\nonumber\\
&&
-16 \Li_3\left(\frac{1-v}{\aj}\right)
-16 \Li_3\left(\frac{\aj}{\aj-v+1}\right)
+8 \Li_3\left(\frac{\aj}{v-1}\right)
-16 \Li_3\left(\frac{v-1}{\aj}\right)
+8 \Li_3\left(-\frac{\aj}{v+1}\right)
\nonumber\\
&&
-8 \Li_3\left(-\frac{2 v}{\aj-v+1}\right)
-16 \Li_3\left(\frac{\aj}{v+1}\right)
+8 \Li_3\left(x^2\right)
-16 \Li_3\left(-\frac{v+1}{\aj}\right)
\nonumber\\
&&
-16 \Li_3\left(\frac{v-1}{-\aj+v-1}\right)
-16 \Li_3\left(\frac{\aj}{\aj+v+1}\right)
-8 \Li_3\left(\frac{2 v}{\aj+v+1}\right)
\nonumber\\
&&
-8 \Li_3\left(\frac{2 \aj v}{(v-1) (\aj+v+1)}\right)
-16 \Li_3\left(\frac{v+1}{\aj+v+1}\right)
+8 F_c \left(\frac{\aj}{\aj-v+1},\frac{\aj}{\aj+v+1}\right)
\Big)
\Big]\, . 
\end{eqnarray*}
The polynomial $Q_S$, the ``conformal" variable $x$ and relative
velocity of the quark pair $v$ read:
\begin{eqnarray}
Q_S &=&  16\left( m_j^2 \piq^2 - 2 \pipj \piq \pjq + m_i^2 \pjq^2 \right) \, , \nonumber\\
x &=& \sqrt{(1-v)/(1+v)} \, ,\nonumber\\
v &=& \sqrt{1-{m_i^2m_j^2\over \pipj^2}} \, . 
\label{eq:variables}
\end{eqnarray}
Beyond order ${\cal O}(\eps^0)$, the result for the one-loop soft
current cannot be expressed in terms of standard
polylogarithms. Multiple polylogarithms appear as evident from
Eq.~(\ref{eq:F1-expand}). At order ${\cal O}(\eps^1)$ we have combined
all functions that are outside the class of the standard
polylogarithms into the function:
\begin{equation}
F_c(x_1,x_2) = \int_0^1 dt  {\ln(1 - t) \ln\left(1 - t{x_2\over x_1}\right) \over {1\over x_2} - t} \, ,
\end{equation}
which can be expressed in terms of multiple polylogarithms of weight
3, see Eq.~(B.22) in Ref.~\cite{Korner:2004rr}. The result
Eq.~(\ref{eq:Result-Case-3}) is also available in electronic form.


\subsection{Properties and checks}\label{sec:checks}

The purpose of the overall $d$-dimensional prefactor in
Eqns.~(\ref{eq:gij-explicitCase1},\ref{eq:gij-explicitCase3}) is to
extract exactly the leading power scaling behavior of the one-loop
soft-gluon current in the limit $q\to 0$. The remainder is given as
expansion in $\eps$, which has a well defined limit $q\to 0$. This is
easy to see since it is invariant under independent rescaling of the
momenta $q,p_i,p_j$.

The result for the one-loop soft-gluon current satisfies a number of
consistency checks. Eq.~(\ref{eq:gij-explicitCase3}) has a well defined
limit, when either one of the masses $m_i$ or $m_j$ vanishes. In the
limit $m_j\to 0$, it agrees with the result for the soft current in
$Case~2$, as it should. Note that this agreement is a non-trivial
check on the analytical continuation used to derive the result in
$Case~2$ from that in $Case~1$. We have numerically checked the result
for the hardest integral $M_3$ in the ``time-like'' kinematics
$Case~3$ (see \ref{sec:app-integrals}).

We have verified that the soft current has the correct behavior in the
small mass limit (see \ref{sec:app-small-mass} for details). The
massless limit $m_i=0,~m_j=0$ of the one-loop un-renormalized soft
current is regular, and the results for the soft current in all
kinematical regions reproduce the massless results of
Ref.~\cite{Catani:2000pi}. 

We have also verified that the pole terms of the one-loop soft current
agree with what is expected based on the structure of the
singularities of massive gauge theory amplitudes (see
\ref{sec:app-poles}). 


\section{Squared matrix elements}\label{sec:squared}

The knowledge of the soft-gluon current makes it possible to construct
an approximation to the squared one-loop matrix element for any
process in the limit (\ref{eq:softlimit}). As indicated in
Eq.~(\ref{eq:M-fact}), this approximation is correct up to power
suppressed terms. The result (\ref{eq:J0}) for the tree-level current
is exact in $\epsilon$. The one-loop current  (\ref{eq:J1}) is
calculated as an expansion in $\epsilon$ which is deep enough to allow
the derivation of the terms ${\cal O}(\epsilon^0)$ in any observable
at NNLO.

In the limit (\ref{eq:softlimit}) the square of a Born amplitude reads:
\begin{eqnarray}
&&\langle M_a^{(0)}(n+1;q) \vert M_a^{(0)}(n+1;q)  \rangle = \nonumber\\
&& -4\pi \alpha_S\mu^{2\epsilon} \Bigg\{ \sum_{i\neq j = 1}^n e_{ij} \langle M^{(0)}(n) \vert T_i\cdot T_j \vert M^{(0)}(n)  \rangle~ +~ \sum_{i = 1}^n {\cal C}_ie_{ii} \langle M^{(0)}(n) \vert M^{(0)}(n)  \rangle \Bigg\} +{\cal O}(\lambda)\, .
\label{eq:M0-square}
\end{eqnarray}
Above we introduced $e_{ij}\equiv e_i\cdot e_j$ and ${\cal C}_i \equiv
T_i\cdot T_i$ is the quadratic Casimir appropriate for the parton
$i$.

The interference term between the Born and one-loop amplitude in the
limit (\ref{eq:softlimit}) reads:
\begin{eqnarray}
&&\langle M_a^{(0)}(n+1;q) \vert M_a^{(1)}(n+1;q)  \rangle  + c.c. = -4\pi \alpha_S\mu^{2\epsilon} \Bigg\{  \nonumber\\
&&  2C_A\sum_{i\neq j = 1}^n \left(e_{ij}-e_{ii}\right) R_{ij}\langle M^{(0)}(n) \vert T_i\cdot T_j \vert M^{(0)}(n)  \rangle  -  4\pi\sum_{i\neq j\neq k = 1}^n e_{ik}I_{ij} \langle M^{(0)}(n)\vert f^{abc}T_i^aT_j^bT_k^c \vert M^{(0)}(n)  \rangle \nonumber\\
&&+ \left( \sum_{i\neq j = 1}^n e_{ij} \langle M^{(0)}(n) \vert T_i\cdot T_j \vert M^{(1)}(n)  \rangle + c.c. \right) + \left( \sum_{i = 1}^n {\cal C}_ie_{ii} \langle M^{(0)}(n) \vert M^{(1)}(n)  \rangle + c.c. \right) \Bigg\} +{\cal O}(\lambda) \, ,
\label{eq:M1-square}
\end{eqnarray}
where we have split $g^{(1)}_{ij} \equiv R_{ij} + i\pi I_{ij}$ into
its real and imaginary parts to be found in
Eqns.~(\ref{eq:gij-explicitCase1},\ref{eq:gij-explicitCase3}).


\section{UV renormalization}\label{sec:UVren}

Up to here we considered bare amplitudes. In practical applications
one works with UV renormalized amplitudes. The UV renormalized
one-loop soft-gluon current is very easy to derive. One needs to
recognize  that though that loop order no mass renormalization
enters. Therefore, all one needs to do is coupling and field
renormalization. Since we consistently set to zero scaleless
integrals, the only correction one needs to take into account is
self-energy contribution in the soft-gluon leg due to the massive
flavors. It will be most convenient to work in a scheme where all
massive flavors are decoupled, i.e. the coupling is running with $n_f$
light flavors only. Then the heavy quark loop contributions into the
external gluon leg will be canceled by the decoupling correction. 
Therefore, in order to obtain the UV renormalized current from the bare one, 
one only needs to perform coupling renormalization 
$\alpha_S^b S_\eps = \alpha_S( 1-\beta_0\alpha_s/(2\pi\eps) + {\cal O}(\alpha_S^2))$ 
in the first line of Eq.(\ref{eq:expansion}). This procedure amounts simply to 
adding the term $\sim\beta_0$ in Eq.~(\ref{eq:J1-poles}) to the bare current; 
see also the discussion following Eq.~(\ref{eq:J1-poles}).


\section{Summary}

In this paper, we have studied the behavior of one-loop QCD amplitudes
with an arbitrary number of massive fermions in the limit when one external
gluon becomes soft. Similarly to the well known massless case, we find
that in the limit (\ref{eq:softlimit}), any amplitude factorizes, up to
power suppressed terms, into a product of a simpler amplitude
and a process independent function: the soft-gluon current.

We have explicitly calculated this current through one loop. This
result enters the evaluation of any cross-section with massive
fermions at next-to-next-to-leading order within a subtraction
approach. An immediate application for this result is the calculation
of the $t{\bar t}$ cross-section at the second perturbative order.

We have performed a number of non-trivial checks on our results. We
have verified that they correctly reproduce their small-mass limit and
pole terms, that we independently predict. We have performed a number
of numerical checks on the non-trivial integrals.

The explicit result for the one-loop soft-gluon current with massive
fermions is much more complicated than in the massless case. While it
is possible to derive an exact result valid in $d$-dimensions, we have
explicitly presented the final result in a form suitable for practical
applications: we have extracted the current's leading behavior exactly
in $d$-dimensions, and expanded the rest in $\eps$ in a form
appropriate for calculating observables at next-to-next-to leading
order. The functional form of the result is significantly more
complicated and involves multiple (Goncharov) polylogarithms.
 
As a by-product of our calculations, and as an additional crosscheck,
we have worked out the analytical continuation from space-like to
time-like kinematics for a multiscale problem. In higher orders in
$\epsilon$ this procedure involves multiple polylogarithms and opens
an interesting subject for further investigation.

\noindent
\section*{Acknowledgments}
M. C. was supported by the Heisenberg and by the Gottfried Wilhelm
Leibniz Programmes of the Deutsche Forschungsgemeinschaft.

\appendix


\section{The scalar integrals}\label{sec:app-integrals}

The one-loop soft-gluon current can be expressed through the following
integrals:
\begin{eqnarray}
M_1 &\equiv& \Phi \int{d^dk\over i(2\pi)^d}{1\over [k^2][(k+q)^2][-p_j\cdot k]} \nonumber\\ 
M_2 &\equiv& \Phi\int{d^dk\over i(2\pi)^d}{1\over [k^2][p_i\cdot k + p_i\cdot q][-p_j\cdot k]} 
\label{eq:masters}\\ 
M_3 &\equiv& \Phi\int{d^dk\over i(2\pi)^d}{1\over [k^2][(k+q)^2][p_i\cdot k + p_i\cdot q][-p_j\cdot k]} 
\nonumber\, ,
\end{eqnarray}
where each propagator has an implicit $+i\delta$ imaginary part. The
momenta $p_i, p_j$ can be massive or massless and the momentum $q$,
corresponding to the soft-gluon, is assumed outgoing and massless. The
normalization factor is $\Phi = 8\pi^2 (4\pi)^{-\eps}e^{\eps \gamma_E}$.

The simplest integral to evaluate is $M_1$:
\begin{eqnarray}
M_1 &=& \Phi~ {\pi^{-2+\eps}\over 16}\Gamma(-\eps)\Gamma(2\eps){m_j^{2\eps}\over \left[ -(p_j\cdot q) - i\delta\right]^{1+2\eps}} \, .
\label{eq:M1}
\end{eqnarray}

Next we consider the integral $M_2$. Its full $q$ dependence can be
extracted, and the remainder expressed through a one-dimensional
integral:
\begin{eqnarray}
M_2 &=&  - \Phi~{\pi^{-2+\eps}\over 4}\Gamma(1-\eps)\Gamma(2\eps) \left[ -(p_i\cdot q) - i\delta\right]^{-2\eps}\nonumber\\
&\times& \int_0^1dt\, t^{-2\eps} \Bigg\{ t^2m_i^2 + (1-t)^2m_j^2 - 2t(1-t)(p_i\cdot p_j)  - i\delta \Bigg\}^{-1+\eps} \, .
\label{eq:M2}
\end{eqnarray}
This one-dimensional integral can be evaluated in terms of
$_2F_1$-type hypergeometric functions. After some rearrangements and
using standard relations between the hypergeometric functions we
obtain:
\begin{eqnarray}
M_2 &=&  \Phi~{\pi^{-2+\eps}\over 4}\Gamma(-\eps)\Gamma(2\eps) \left[ -(p_i\cdot q) - i\delta\right]^{-2\eps} \left[ -2(p_i\cdot p_j) - i\delta\right]^{-1+\eps}\nonumber\\ 
&\times& \Bigg\{{\Gamma(1+\eps)\Gamma(1-2\eps) \over \Gamma(1-\eps) }~v^{-1+2\eps}~ \beta_j^{-\eps} - {2 \, \beta_i^{\eps} \over 1+v}\, {_2F_1}\left(1,1-\eps,1+\eps;{1-v\over 1+v}\right)\Bigg\} \, . 
\label{eq:M2-exact}
\end{eqnarray}
We have introduced $\beta_k \equiv m_k^2/(-2(p_i\cdot p_j) -
i\delta),~k=i,j$, and the relative velocity $v$ defined in
Eq.~(\ref{eq:variables}). The hypergeometric function can be expanded
in series in $\eps$ to any desired depth with the help of
\cite{Huber:2005yg}.

The most complicated integral is $M_3$. We first apply Schwinger
$\alpha$-parameterization. The two integrations, corresponding to the
two propagators quadratic in $k$, can be transformed in the usual way:
\begin{eqnarray}
\int_0^\infty d\hat\alpha_1 d\hat\alpha_2 = \int_0^\infty a da\int_0^1dy \, ,
\label{eq:change-var}
\end{eqnarray}
and the integration over $a$ performed. 

In order to extract the scaling behavior of the integral in the limit
$q\to 0$, we rescale the $\alpha$-parameters $\hat\alpha_{3,4}$
corresponding to the two propagators that are linear in $k$:
\begin{eqnarray}
\hat\alpha_3 \to \hat\alpha_3/\vert p_i\cdot q\vert ~~,~~ \hat\alpha_4 \to \hat\alpha_4/\vert p_j\cdot q\vert \, . 
\end{eqnarray}
We note that the invariants $(p_i\cdot q)$ and $(p_i\cdot p_j)$ are
non-zero, although their signs change depending on the kinematical
configuration. Next we change the variables $\hat\alpha_{3,4}$ along
the lines of Eq.~(\ref{eq:change-var}) and perform the integration
over the infinite range, arriving at the following two-dimensional
representation for $M_3$:
\begin{eqnarray}
M_3 &=& \Phi~{\pi^{-2+\epsilon}\over 16}\Gamma(-\eps)\Gamma(2+2\eps) {1\over \vert p_i\cdot q\vert\vert p_j\cdot   q\vert} \int_0^1dt\int_0^1dy \left( t^2{m_i^2\over (p_i\cdot q)^2} + (1-t)^2{m_j^2\over (p_j\cdot q)^2} \right. \nonumber\\ 
&& \left. - 2t(1-t){(p_i\cdot p_j)\over \vert p_i\cdot q\vert\vert p_j\cdot   q\vert}  - i\delta \right)^\eps \left( - y\sigma_j-t\sigma_i +(\sigma_i+\sigma_j)ty - i\delta \right)^{-2-2\eps} \, ,
\label{eq:M3-2dim}
\end{eqnarray}
where $\sigma_k\equiv (p_k\cdot q)/\vert p_k\cdot q\vert = \pm
1$. Note that the signs $\sigma_{i,j}$ also determine the sign of
$(p_i\cdot p_j)$.

The evaluation of Eq.~(\ref{eq:M3-2dim}) is hardest in the
phenomenologically relevant ``time-like'' (TL) kinematics of {\it Case
  3}, where $\sigma_i= \sigma_j =1$. Performing the $y$-integration we
get:
\begin{eqnarray}
&&M_3^{(TL)} = \Phi~{\pi^{-2+\epsilon}\over 16(p_i\cdot p_j)}\Gamma(-\eps)\Gamma(1+2\eps) \left(-1-i\delta\right)^{-2\epsilon}\left({p_i\cdot p_j\over (p_i\cdot q)(p_j\cdot   q)}\right)^{1+\eps} \int_0^1dt \left( t^2{m_i^2(p_j\cdot q)\over (p_i\cdot q)(p_i\cdot p_j)}\right. \nonumber\\
&& \left.+ (1-t)^2{m_j^2(p_i\cdot q)\over (p_j\cdot q)(p_i\cdot p_j)} - 2t(1-t) -i\delta \right)^\eps \left[ {t^{-1-2\eps} - (1-t)^{-1-2\eps} \over 1-2t}\right] \, .
\end{eqnarray}
We evaluate the above integral as expansion in $\epsilon$ using the
formula:
$$z^{-1+\eps} = {1\over \eps}\delta(z) + \sum_{k=0}^\infty {\eps^n\over k!} \left( {\ln(z)\over z}\right)_+ \, .$$
Extracting the imaginary parts is particularly laborious since both
roots $x_{1,2}^t$ of the polynomial, which is quadratic in $t$, are inside the
integration range: $0\leq x_2^t \leq 1/2 \leq x_1^t \leq 1$. The
expressions for the two roots in the time-like case read:
\begin{equation}
x_1^t = {\alpha_j\over \alpha_j + 1-v } ~,~  x_2^t = {\alpha_j\over \alpha_j+1+v} \, ,
\end{equation}
with $\ai,\aj$ defined in Eq.~(\ref{eq:def-ai-aj}). The explicit
result for the integral $M_3^{(TL)}$ is rather lengthy. It is supplied
in electronic form.

The integral $M_3$ in Eq.~(\ref{eq:M3-2dim}) is much easier to
calculate in the ``space-like'' (SL) kinematics $\sigma_i= - \sigma_j
=1$; in the following we present its derivation for
completeness. Performing the $y$-integration and after some
simplifications we obtain:
\begin{eqnarray}
&&M_3^{(SL)} = \Phi~{\pi^{-2+\epsilon}\over 16(p_i\cdot p_j)}\Gamma(-\eps)\Gamma(1+2\eps) \left({p_i\cdot p_j\over (p_i\cdot q)(p_j\cdot   q)}\right)^{1+\eps} \int_0^1dt \left( t^2{m_i^2(p_j\cdot q)\over (p_i\cdot q)(p_i\cdot p_j)}\right. \nonumber\\
&& \left.+ (1-t)^2{m_j^2(p_i\cdot q)\over (p_j\cdot q)(p_i\cdot p_j)} + 2t(1-t)  \right)^\eps \left[ (1-t)^{-1-2\eps} + (-1-i\delta)^{-2\eps}t^{-1-2\eps}\right] \, .
\label{eq:M3-1dim}
\end{eqnarray}
The one-dimensional integral can be evaluated in terms of the Appell
hypergeometric function $F_1$:
\begin{eqnarray}
&&M_3^{(SL)} = - \Phi~{\pi^{-2+\epsilon}\over 16}{\Gamma(-\eps)\Gamma(2\eps)\over (p_i\cdot q)(p_j\cdot q)} \left({p_i\cdot p_j\over (p_i\cdot q)(p_j\cdot   q)}\right)^{\eps} \Bigg\{  \label{eq:J4-1-dim}\\
&&\left(-1-i\delta\right)^{-2\eps} \alpha_j^\eps F_1\left( -2\eps,-\eps,-\eps,1-2\eps; {1\over x_1^{s}},{1\over x_2^{s}}\right)  +  \alpha_i^\eps F_1\left( -2\eps,-\eps,-\eps,1-2\eps; {1\over 1-x_2^{s}},{1\over 1- x_1^{s}}\right) \Bigg\} \, .\nonumber
\end{eqnarray}
In the above equation we have introduced the following notation:
\begin{equation}
x_1^s = {\alpha_j\over \alpha_j-1-v} ~,~  x_2^s = {\alpha_j\over \alpha_j-1+v} \, .
\label{eq:SL-roots}
\end{equation}
The quantities $x_{1,2}^s$ are the two roots of the
polynomial quadratic in $t$ appearing in Eq.~(\ref{eq:M3-1dim}) with
$x_1^s<0$ and $x_2^s>1$ and $\ai,\aj$ are defined in
Eq.~(\ref{eq:def-ai-aj}).

The Appell functions can be expanded in $\eps$ in terms of multiple
polylogarithms $\Li_{m_k,\dots,m_1}(t_k,\dots , t_1)$ with the help of
the library {\it Nestedsums}~\cite{nestedsums} (see also
\cite{Moch:2005uc}):
\begin{eqnarray}
&& F_1(-2\eps,-\eps,-\eps,1-2\eps;t,y) = 1 + \eps^2\left[ 2\Li_2(t) + 2\Li_2(y)  \right] \nonumber\\
&& + \eps^3\left[ 4\Li_3(t) + 4\Li_3(y) - 2S_{1,2}(t) -2S_{1,2}(y) - 2\Li_{1,2}\left({t\over y},y\right) - 2\Li_{1,2}\left({y\over t}, t\right)  \right] \nonumber\\
&&+ \eps^4\left[ 8\Li_4(t) + 8\Li_4(y) - 4S_{2,2}(t) -4S_{2,2}(y) + 2S_{1,3}(t) +2S_{1,3}(y) - 4\Li_{1,3}\left({t\over y},y\right) - 4\Li_{1,3}\left({y\over t}, t\right) \right.\nonumber\\
&&+2\Li_{1,1,2}\left(1,{t\over y},y\right)+2\Li_{1,1,2}\left(1,{y\over t},t\right) +2\Li_{1,1,2}\left({t\over y},1,y\right)+2\Li_{1,1,2}\left({y\over t},1,t\right) \nonumber\\
&&\left.+2\Li_{1,1,2}\left({y\over t},{t\over y},y\right)+2\Li_{1,1,2}\left({t\over y},{y\over t},t\right) \right]  + {\cal O}(\eps^5) \, .
\label{eq:F1-expand}
\end{eqnarray}
The functions $\Li_n(t)$ are the usual polylogarithms and $S_{n,p}(t)$
are the Nielsen's generalized polylogarithms. We follow the
conventions and definitions of Ref.~\cite{nestedsums} (see also
Ref.~\cite{Moch:2001zr}). The numerical evaluation of multiple
polylogarithms has been automated in Ref.~\cite{Vollinga:2004sn}.

The results given above are sufficient to explicitly derive the
one-loop soft-gluon current in the kinematics where one of the massive
quarks is in the initial state. Such formal result is of interest, for
example, in studies of the properties of massive gauge-theory
amplitudes.

In the case of one non-zero mass, as needed for {\it Case 1}, the
integral $M_3^{(SL)}$ reads:
\begin{eqnarray}\label{eq:M3SL}
&&M_3^{(SL)}\vert_{m_j=0} = \Phi~{\pi^{-2+\epsilon}\over 16}{\Gamma(-\eps)\Gamma(1+2\eps)\over (p_i\cdot q)(p_j\cdot q)} \left({2\, p_i\cdot p_j\over (p_i\cdot q)(p_j\cdot   q)}\right)^{\eps} \Bigg\{  \label{eq:M3-1mass}\\
&&  {\Gamma(1+\eps)\Gamma(-2\eps)\over \Gamma(1-\eps)} {_2F_1}\left( -\eps,1+\eps,1-\eps; 1-{\alpha_i\over 2}\right) -{ \left(-1-i\delta\right)^{-2\eps}\over \eps}~ {_2F_1}\left( -\eps,-\eps,1-\eps; 1-{\alpha_i\over 2}\right)  \Bigg\} \, .\nonumber
\end{eqnarray}

In the massless case, the result is simple and agrees with the one
given in Ref.~\cite{Catani:2000pi}.

Finally, we remark that even in the case of equal masses, the result
for the one-loop soft-gluon current is a function of two independent
parameters, i.e. the case of two un-equal masses is not more
complicated than the equal mass case. This is evident, for example,
from Eq.~(\ref{eq:J4-1-dim}).


\section{Small-mass limit of $ J_a^{\mu (1)} $}\label{sec:app-small-mass}

Following the methods of Ref.~\cite{Mitov:2006xs} one can
independently derive the leading behavior of the one-loop soft-gluon
current in the small-mass limit. Considering UV renormalized
amplitudes, assuming all non-zero masses are equal and then taking the
small-mass limit of both sides in Eq.~(\ref{eq:M-fact}), one easily
derives:
\begin{equation}
J_a(q;m\neq 0) = \sqrt{Z_{[g]}^{(m|0)}} J_a(q;m= 0) + {\cal O}(m^2) \, .
\label{eq:app-massless-nh}
\end{equation}
In the above equation we take the strong coupling running with $n_L+1$
flavors, i.e. the heavy flavor is active. The factor $Z_{[g]}^{(m|0)}$
is given in the appendix of Ref.~\cite{Mitov:2006xs} through one loop
and to all orders in $\eps$. One can also check that upon decoupling
the heavy flavor (in $d$-dimensions) the $Z$-factor in the above
equation is exactly compensated through one loop, i.e. in the
decoupling scheme, Eq.~(\ref{eq:app-massless-nh}) simplifies to:
\begin{equation}
J_a(q;m\neq 0) =  J_a(q;m= 0) + {\cal O}(m^2) \, .
\label{eq:app-massless-nl}
\end{equation}

The decoupling relations in $d$-dimension can be found, for example,
in Ref.~\cite{Gluza:2009yy}; see also Section~\ref{sec:UVren}.


\section{Pole structure of $ J_a^{\mu (1)} $}\label{sec:app-poles}

One can provide an independent derivation of the poles of the UV
renormalized, one-loop soft-gluon current from the known structure of
the singularities of one-loop gauge theory amplitudes. Considering the
amplitudes appearing in Eq.~(\ref{eq:M-fact}) as wide-angle scattering
amplitudes, we can decompose them into jet, soft and hard functions
respectively \cite{Sen:1982bt}:
\begin{eqnarray}
M_a(n+1;q) &=& I \times S_{ab} \cdot H_b \, , \nonumber\\
M(n) &=& i \times \sigma\cdot h \, .
\label{eq:JSH}
\end{eqnarray}
The jet functions $I,i$ are diagonal in color. At one-loop they
contain double and single poles. The soft functions $S,\sigma$ are
color matrices that have single poles only. The hard functions $H,h$
are finite color vectors.

Applying the decomposition (\ref{eq:JSH}) to Eq.~(\ref{eq:M-fact}) and
expanding each factor through one loop we get:
\begin{eqnarray}
J_a^{(1)} &=& \left( I^{(1)}-i^{(1)} - \sigma^{(1)}\right) J_a^{(0)} + S^{(1)}_{ab} J_b^{(0)} + \left[\sigma^{(1)},J_a^{(0)}\right] + {\cal O}(\eps^0)\, .
\end{eqnarray}

The separation of jet and soft functions is scheme dependent. We work
in the formfactor scheme \cite{Sterman:2002qn} where the jet function
is a product of the square root of the formfactors corresponding to
each external leg, and similarly in the massive case
\cite{Mitov:2006xs}. It then immediately follows that $I^{(1)}-i^{(1)}
= f_g^{(1)}/2$, where $f_g^{(1)}$ is the one-loop correction to the
UV-renormalized gluon form-factor
\cite{Baikov:2009bg,Gehrmann:2010ue}:
\begin{equation}
f_g^{(1)} = {\alpha_S\over 2\pi} \left( -{C_A\over \eps^2}-{\beta_0\over\eps}+{\cal O}(\eps^0) \right) \, ,
\end{equation}
where $\beta_0 =11C_A/6 - N_F/3$ and $\alpha_S$ is the ${\overline
  {\rm MS}}$ renormalized coupling at scale $\mu$.

The explicit results for the soft functions read:
\begin{eqnarray}
\sigma^{(1)} &=& {\alpha_S\over 2\pi\eps}~{1\over 2}\sum_{i\neq j=1}^n s_{ij}T_i\cdot T_j \, ,\nonumber \\
S^{(1)}_{ab} &=& \sigma^{(1)}\delta_{ab} + {\alpha_S\over 2\pi\eps}~ \sum_{i=1}^n s_{gi}\left(T_g\right)_{ab}\cdot T_i \, ,
\end{eqnarray}
and the index $g$ denotes the soft-gluon leg. The color matrices
pertaining to the soft-gluon are $(T_g^a)_{cb} = if_{cab}$. The
functions $s_{ij}=s_{ji}$, not to be confused with partonic
invariants,
\footnote{In this paper we do not use the notation $s_{ij}$ to denote
  partonic invariants.}
depend on whether the legs $i,j$ are both massive or not and can be
found, for example, in Eq.~(29) of Ref.~\cite{Czakon:2009zw}.

Combining the above results and after some algebra we derive the
following expression for the poles of the UV renormalized one-loop
soft-gluon current:
\begin{eqnarray}
J_a^{\mu (1)} = {\alpha_S\over 2\pi} ~ if^{abc}\sum_{i\neq j=1}^n T_i^bT_j^c \left(e_i^\mu-e_j^\mu\right) \Bigg\{ -{1\over 2\eps^2} -{1\over 2\eps}\left[  {\beta_0\over C_A} + \ln\left({-\mu^2(p_i\cdot p_j)\over 2(p_i\cdot q)(p_j\cdot q)}\right) - h_{ij}\right] + {\cal O}(\eps^0) \Bigg\}\, .
\label{eq:J1-poles}
\end{eqnarray}
The function $h_{ij}$ reads (see, for example, Eq.~(29) of
Ref.~\cite{Czakon:2009zw}):
\begin{equation}
h_{ij} = \ln\left(1+x^2\right) + {2x^2\over 1-x^2}\ln(x) \, , ~ ~{\rm when} ~~ m_i\neq 0, m_j\neq 0 \, ,
\end{equation}
and zero otherwise. The variable $x$ is defined in
Eq.~(\ref{eq:variables}). The function $h_{ij}$ vanishes in the
massless limit. In deriving the above result we have used color
conservation $\sum_{i=1}^n T_i^a=0$ and the identity
$if^{abc}T_i^aT_i^b=-(C_A/2) T_i^c$, or alternatively, $ J_a^{\mu (0)}
= {if^{abc}\over C_A} \sum_{i\neq j=1}^n T_i^bT_j^c
\left(e_i^\mu-e_j^\mu\right)$.

Eq.~(\ref{eq:J1-poles}) agrees with Ref.~\cite{Catani:2000pi}. To that
end we need to convert the renormalized coupling to the bare one and 
recall the overall factor of $g_S$ in Eq.~(\ref{eq:expansion}). 
We also recall the discussion in Section \ref{sec:UVren} where we explain 
that we work with $n_f=n_l$ active flavors and that heavy quark loops in 
external gluon fields are exactly compensated by the decoupling relation. 
See also Refs.~\cite{Becher:2007cu},\cite{Czakon:2007wk}, for more information on that point. 

Eq.~(\ref{eq:J1-poles}) applies to space-like kinematics as in {\it Case 1}. 
Continuation to any other kinematics is trivial; see Ref.~\cite{Mitov:2010xw} and \ref{sec:analytical-cont}.


\section{Analytical continuation to physical kinematics}\label{sec:analytical-cont}

It is often the case that calculations of scattering amplitudes are
easier to perform in unphysical kinematics. Then the question arises
how to analytically continue the result derived in such unphysical
kinematics to the physical region. In this work, the continuation
involves the momentum $p_j$, i.e. we need to continue results derived
in kinematics where $p_j$ is incoming to kinematics where $p_j$ is
outgoing.

The analytical continuation is more involved when $p_j$ is
massive. When the momentum $p_j$ is incoming we have evaluated the
soft current exactly in $d$-dimensions (see
Eqns.~(\ref{eq:gij-masters}, \ref{eq:M1}, \ref{eq:M2-exact},
\ref{eq:J4-1-dim}, \ref{eq:F1-expand}, \ref{eq:M3SL}). To continue the result to the
timelike kinematics where $p_j$ is outgoing one has to first express
the result in a minimal number of variables. Given the
scaling-invariance properties of the result, only two variables are
truly independent. As such we take $x$ and $\ai$, defined in
Eqns.~(\ref{eq:def-ai-aj},\ref{eq:variables}). The variables $\aj$ and
$v$ can be eliminated through the relations $\ai\aj = 1-v^2$ and
$v=(1-x^2)/(1+x^2)$.

The rules for the analytical continuation in terms of the variables
$x$ and $\ai$ (for general values of the masses $m_{i,j}$) are simple, see also Ref.~\cite{Mitov:2010xw}:
\begin{eqnarray}
&& x \to - x + i\delta \, ,\nonumber\\
&& \ai \to \ai \, .
\label{eq:an-cont}
\end{eqnarray}
The invariance of $\ai$ is easy to understand, since $\ai \sim
\pjq/\pipj$. Not only is $\ai$ invariant under $p_j\to - p_j$
\footnote{Strictly speaking one inverts not the momentum $p_j$ but the
  sign of all invarians $(k\cdot p_j)$ linear in $p_j$.}
but, more importantly, its log is: $\ln(\ai) = \ln(\pjq) - \ln(\pipj)
+\dots = {\rm inv}$. That is distinct from the case of $\aj\sim
1/(\pjq\pipj)$ which itself is invariant but its log is not: $\ln(\aj)
= -\ln(\pjq)-\ln(\pipj) +\dots \neq {\rm inv}$. That $\aj$ must
transform nontrivially also follows from the identity $\ai\aj=1-v^2$.

The practical implementation of the analytical continuation procedure
requires the explicit extraction of all branching-point singularities
around the point $x = 0$. Here is a typical example:
\begin{equation}
\Li_2\left(-{1\over x^2}\right)  =  - \Li_2(-x^2) - {\pi^2\over 6} - 2\ln^2(x) \, ,
\end{equation}
and:
\begin{equation}
\ln(x^2) \to \ln(x^2) + 2i\pi ~~;~~ \Li_n(1 - x^2) \to \Li_n(1 - x^2) - 2i\pi {\ln^{n - 1}(1 - x^2)\over (n - 1)!}\, .
\end{equation}

We have verified that with the help of Eq.~(\ref{eq:an-cont}) we can
reproduce the first three orders in $\eps$ of the directly calculated
integral $M_3^{(TL)}$, from the spacelike calculation of $M_3^{(SL)}$;
see \ref{sec:app-integrals}. Starting from the fourth order in $\eps$
one would have to devise a similar procedure for the set of multiple
polylogarithms that begin to appear, see
Eq.~(\ref{eq:F1-expand}). This presents an interesting direction for
future work that can benefit from the number of recent applications of
this class of functions in the context of gauge amplitudes in ${\cal
  N}=4$ SYM theories \cite{Goncharov:2010jf}.

When $p_j$ is massless, the analytical continuation allows one to
obtain the results for configuration $Case~2$ from the one for
$Case~1$. For $m_j=0$ the conformal variable vanishes, $x=0$, which
implies that the analytical continuation (\ref{eq:an-cont}) becomes
trivial. That can also be seen with a direct inspection of the
integrals (\ref{eq:M1},\ref{eq:M2}) and (\ref{eq:M3-2dim}) in the case
$m_j=0$.


\end{document}